\newif\ifsingle

\ifsingle
\documentclass[11pt,draftclsnofoot, onecolumn]{IEEEtran}		
\else		
\documentclass[10pt,final, twocolumn]{IEEEtran}
\fi

\usepackage{times}
\usepackage{amsmath,dsfont}
\usepackage{amssymb,amsthm}
\usepackage{epsfig,verbatim}
\usepackage{subfigure}
\usepackage{setspace}
\usepackage{color}
\usepackage{cite}
\usepackage{epstopdf}
\usepackage{graphics}
\usepackage{accents}
\usepackage{acronym}
\usepackage[bookmarks,colorlinks]{hyperref}
\usepackage{booktabs}
\usepackage{mathtools}
\usepackage{algorithm}
\usepackage{algorithmic}
\usepackage{enumitem}
\usepackage{pstool}
\usepackage{multirow}
\usepackage{soul}

\ifsingle

\else

\fi 

\acrodef{adc}[ADC]{Analog-to-Digital Convertor}
\acrodef{dac}[DAC]{digital-to-analog convertor}
\acrodef{cs}[CS]{Compressed Sensing}
\acrodef{dtft}[DTFT]{discrete-time Fourier transform}
\acrodef{dnn}[DNN]{deep neural network}
\acrodef{csi}[CSI]{channel state information}
\acrodef{map}[MAP]{maximum a-posteriori probability}
\acrodef{snr}[SNR]{signal-to-noise ratio}
\acrodef{sinr}[SINR]{signal-to-interference-and-noise ratio}
\acrodef{bs}[BS]{Base Station}
\acrodef{iot}[IOT]{Interent of Things}
\acrodef{mimo}[MIMO]{Multiple-Input Multiple-Output}
\acrodef{mse}[MSE]{mean-squared error}
\acrodef{pdf}[PDF]{probability density function}
\acrodef{rv}[RV]{random variable}
\acrodef{fec}[FEC]{forward error correction}
\acrodef{rs}[RS]{Reed-Solomon}
\acrodef{lti}[LTI]{linear time-invariant}
\acrodef{wss}[WSS]{wide-sense stationary}
\acrodef{psd}[PSD]{power spectral density}
\acrodef{ser}[SER]{symbol error rate}
\acrodef{ber}[BER]{bit error rate}
\acrodef{isi}[ISI]{intersymbol interference}
\acrodef{awgn}[AWGN]{additive white Gaussian noise}
\acrodef{ut}[UTs]{User Terminals}
\acrodef{mmw}[mmWave]{millimeter wave}
\acrodef{ris}[RIS]{reconfigurable intelligent surface}
\acrodef{dma}[DMA]{Dynamic Metasurface Antenna}
\acrodef{5G}{fifth generation}

\IEEEoverridecommandlockouts

\begin{document}
\title{Hardware Implementation of Task-based Quantization in Multi-user Signal Recovery
}
	\author{
	 Xing Zhang,~\IEEEmembership{Member,~IEEE}, Haiyang Zhang,~\IEEEmembership{Member,~IEEE}, Nimrod Glazer,~\IEEEmembership{Member,~IEEE}, Oded Cohen, Eliya Reznitskiy, Shlomi Savariego, Moshe Namer, and Yonina C. Eldar,~\IEEEmembership{Fellow,~IEEE}	

\thanks{X. Zhang, and H. Zhang are with the School of Communication and Information Engineering, Nanjing University of Posts and Telecommunications, Nanjing, China.
(e-mail: \{20220154, 20220142\}@njupt.edu.cn).
}
\thanks{N. Glazer, O. Cohen, E. Reznitskiy, S. Savariego, and Y. C. Eldar are with the Faculty of Math and Computer Science, Weizmann Institute of Science, Israel. (e-mail:
\{nimrod.glazer, oded.cohen, eliya.reznitskiy, shlomi.savariego, yonina.eldar\}@weizmann.ac.il).}
\thanks{M. Namer is with the EE Department, Technion Israel Institute of Technology, Haifa, Israel
(e-mail: namer@ee.technion.ac.il)}
	
\vspace{-0.75cm}
}
	\maketitle
	\pagestyle{plain}
	\thispagestyle{plain}
	

\begin{abstract}
Quantization plays a critical role in digital signal processing systems, allowing the representation of continuous-amplitude signals with a finite number of bits. However, accurately representing signals requires a large number of quantization bits, which causes severe cost, power consumption, and memory burden. A promising way to address this issue is task-based quantization. By exploiting the task information for the overall system design, task-based quantization can achieve satisfying performance with low quantization costs. In this work, we apply task-based quantization to multi-user signal recovery and present a hardware prototype implementation. The prototype consists of a tailored configurable combining board, and a software-based processing and demonstration system. Through experiments, we verify that
with proper design, the task-based quantization achieves a reduction of $25$ fold in memory by reducing from $16$ receivers with $16$ bits each to $2$ receivers with $5$ bits each, without compromising signal recovery performance.

{\textbf{\textit{Index terms---}}}
Task-based quantization, multi-user signal recovery, analog combiner, hardware implementation.
\end{abstract}
%
\section{Introduction}
\label{sec:Introduction}
Processing and storing information that originates as analog signals involves converting this information to bits by analog-to-digital converters (ADCs)\cite{Eldar_Sampling}. In conventional receivers, the ADC is employed as a separate unit regardless of other parts of the system. ADCs typically sample at the Nyquist rate of the received signal and use high-resolution quantizers, so that sampling and quantization errors can be minimized.
However,
since the power consumption of ADCs and the required storage memory grow with the sampling rate and quantization resolution, conventional ADCs pose great challenges to practical applications with high data rates.
For example,
in future 6G wireless communication systems, where
hundreds or even thousands of antennas and millimeter-wave (mmWave) or sub-terahertz (THz)
signaling are employed,
it is expected that up to 1 Tb/s data rate will be achieved \cite{rajatheva2020white,letaief2019roadmap}. In such cases, the hardware implementation of high-resolution ADCs becomes a bottleneck.
Therefore, more efficient sampling and quantization schemes are necessary.

Two prominent research directions to alleviate the power burden of ADCs are sub-Nyquist sampling and low-resolution quantization.
Sub-Nyquist sampling aims to reduce the sampling rate by exploiting the underlying structure information of the signal \cite{Eldar_Sampling,Kipnis_Analog2018}. For instance, the nature of finite rate of innovation (FRI) signals has been exploited to reduce the sampling rate of received signals in ultrasound \cite{Drori_Compress2021}, radar \cite{Cohen_Sub2018} and cognitive radio\cite{Cohen_Analog2018}.
However, the sub-Nyquist sampling framework does not take into account the effect of quantization.
As the power consumption of ADC increases in an exponential manner with the number of quantization bits, low precision quantization  has attracted great interest in recent years.
Low-resolution quantization uses a few or even 1-bit to
discretize the signal amplitude.
It has been applied in various applications such as massive multi-input multi-output (MIMO) communications \cite{jacobsson2017throughput,dong2020spatially}, radar \cite{ameri2019one}, direction of arrival estimation \cite{sedighi2021doa}, and spectrum sensing \cite{ali2017power}.
Compared with conventional high resolution quantizers, significant rate reduction can be expected by using low-bit quantizers.
However, compensation for the distortion induced in quantization is required in subsequent digital processing, which results in complicated information extraction in the digital domain and an overall system performance degradation.

To address these issues, the authors of \cite{Shlezinger_Tsp19} proposed task-based quantization.
By taking into account the underlying task in the system design, task-based quantization dramatically reduces the number of bits while allowing for accurate signal recovery.
This is achieved by introducing an analog combiner, followed by joint optimization of the analog and digital processing and the bridge between them, i.e., the ADC. Task-based quantization has been applied to graph signal processing \cite{PeiLi_Arx21},
channel estimation in massive MIMO communications \cite{Shlezinger_Tsp19Asym}, and target identification in radar \cite{Feng_Tsp21}. Theoretical maturity of the concept suggests the need to demonstrate and evaluate the implementation of such systems in hardware, which is the focus of this work.

Here,
we apply task-based quantization to multi-user signal recovery and present a prototype, which consists of a hardware board and a software-aided demonstration system. In the considered setting, the system task is to recover multi-user transmitted signals, rather than the received signals on all antennas. Therefore, following the principle of task-based quantization,
a tailored analog combiner board was built to properly pre-process the received signals prior to quantization. The outputs are then quantized by scalar quantizers with limited bits. Finally, the task vector is recovered by an optimized digital matrix.
To visually demonstrate the above process,
a MATLAB-based graphical user interface (GUI) was developed, which includes  parameter controlling, data processing and results displaying.
Experimental results illustrate the superiority of task-based quantization over the conventional task-ignorant one, mitigating the gap between the theory and its practical application.

The rest of this paper is organized as follows: Section \ref{sec:backG} formulates the problem of task-based quantization for multi-user signal recovery and provides the theoretical results. Next, in Section \ref{sec:hardware}, the system architecture and each component of the hardware prototype are introduced in detail. Experimental results are provided in Section \ref{sec:Experimental}, followed by conclusions in Section \ref{sec:Conclusions}.

\textit{Notation:} Scalar quantities, column vectors and matrices are denoted by lowercase letters, $a$, bold lowercase letters, $\textbf{a}$, and bold uppercase letters, $\textbf{A}$, respectively. The superscripts ${{(\cdot)}^{T}}$, ${{(\cdot)}^{-1}}$ and ${{(\cdot)}^{H}}$ are, respectively, the transpose, inverse and Hermitian transpose operators. The symbol $E[\cdot]$ represents statistical expectation, $||\cdot||$ is the Euclidean norm, $\mathcal{C}$ is the set of complex numbers, and $\textbf{I}_K$ is the $K\times K$ identity matrix. We use $a^{+}$ to denote $\text{max}(a,0)$, and $\lfloor\cdot\rfloor$ to denote rounding down to the next smaller integer.
%
\section{Task-based Quantization for Multi-user Signal Recovery}
\label{sec:backG}
\begin{figure}[h]
    \centering
    \includegraphics[width=0.48\textwidth]{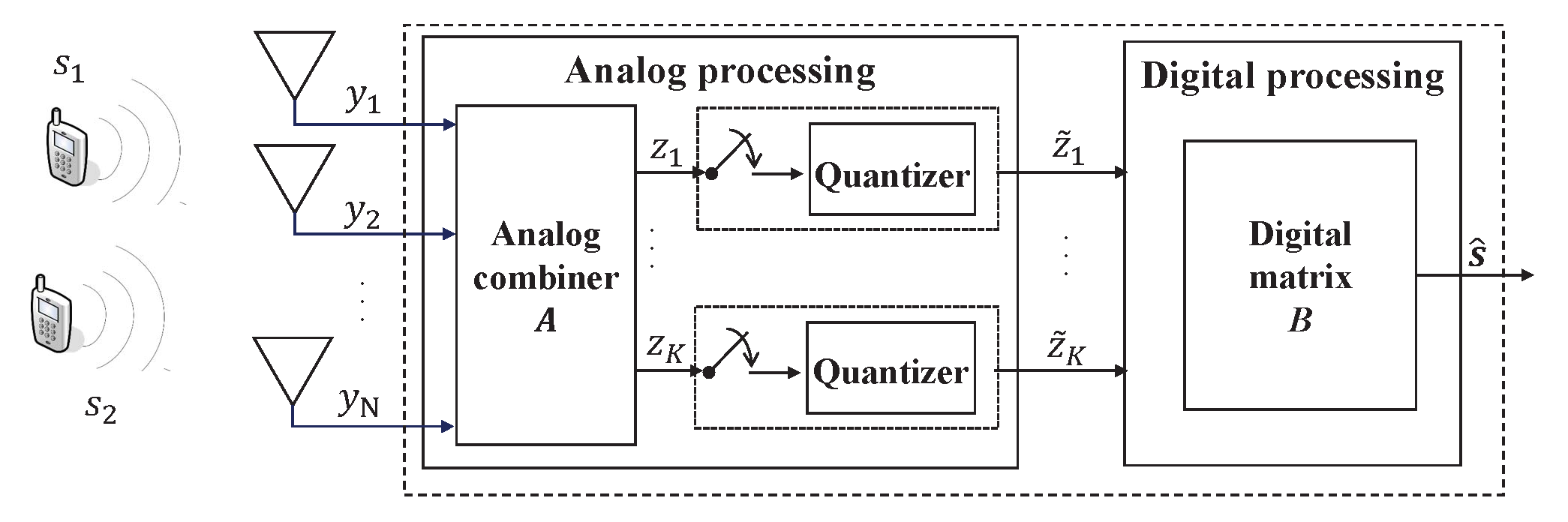}
    \caption{Illustration of task-based quantization for multi-user signal recovery.}
    \label{Fig1}
\end{figure}
In this section, we provide a mathematical description of multi-user signal recovery under task-based quantization. In particular, we begin by introducing the system model of task-based quantization for multi-user signal recovery in Subsection \ref{subsec:SM}, followed by theoretical results in Subsection
\ref{subsec:TR}.
\subsection{System Model}
\label{subsec:SM}
Consider a single-cell network in which a base station (BS) is equipped with $N$ antennas and serves $K$ single-antenna user terminals (UTs), as shown in Fig. \ref{Fig1}. In uplink, let a $K\times 1$ vector $\textbf{s}$ be the transmitted signals of all the UTs in the cell at one time instant. The received $N\times 1$ signal vector $\textbf{y}$ at the BS can be expressed as
\begin{equation}
\textbf{y}=\textbf{Hs}+\textbf{v},
\end{equation}
where $\textbf{v}$ represents additive white Gaussian noise (AWGN).
The $N\times K$ matrix $\textbf{H}$ denotes the wireless channel, with its $k$-th column representing the channel between user $k$ and the antenna array, given by \cite{yang2020uplink,he2019super}
\begin{equation}
\textbf{h}_k=
g_k e^{-j2\pi \frac{f_c r_k}{c}}\textbf{a}(\theta_{k}).
\label{Eq: channel}
\end{equation}
Here $g_k$ denotes the path gain, where without loss of generality, we assume only the line of sight (LoS) path exists, $c$ is the speed of light, $f_c$ is the carrier frequency, $r_k$ and $\theta_k$ are respectively the distance and angle of arrival of the $k$th user. The vector $\textbf{a}(\theta_k)$ is the the steering vector, given by
\begin{equation}
\textbf{a}({\theta_k})=\left[ 1, e^{j\pi \text{sin}\theta_k}, \cdots, e^{j\pi (N-1)\text{sin}\theta_k}\right]^T.
\end{equation}
We assume the channel is quasi-static over the signal transmission time, so that $\textbf{H}$ can be estimated by pilots and is assumed to be known for the task of recovering $\textbf{s}$.

In conventional quantization systems, ADCs are only used to discretize the received signals. The task of recovering $\textbf{s}$ is performed separately in the digital domain. By contrast, task-based quantization proposed in \cite{Shlezinger_Tsp19} jointly designs the overall analog and digital system to estimate $\textbf{s}$.
Specifically, the received signal $\textbf{y}$ is first projected to a $K\times 1$ vector $\textbf{z}$ by using an analog combiner $\textbf{A}$, i.e.,
\begin{equation}
\textbf{z}=\textbf{Ay}=\textbf{AHs}+\textbf{Av}.
\end{equation}
Then, each entry of $\textbf{z}$ is sampled and  quantized using scalar quantizers with dynamic range $\gamma$ and resolution $\tilde{M}_K \triangleq \lfloor M^{1/K} \rfloor$.
The symbol $M$ is the overall number of quantization levels, which represents the memory requirement of the system and is also directly related to the ADC power consumption.
When the input is inside the dynamic range of the quantizer, the output can be written as the sum of the input and an additive zero-mean white noise signal according to the theory of dithered quantization \cite{Gray_Tit93Dither}, that is,
\begin{equation}
\tilde{\textbf{z}}=\textbf{AHs}+\textbf{Av}+\textbf{e},
\end{equation}
where $\textbf{e}$ is the quantization noise with covariance $\frac{\Delta^2}{2}\textbf{I}_K$. The symbol $\Delta$ denotes the quantization spacing defined as $\Delta=\frac{2\gamma}{\tilde{M}_K}$. When $\tilde{M}_K$ is given, the value of $\gamma$ determines the quantization spacing, and therefore, the variance of the quantization noise.

In the digital domain, the estimation of $\textbf{s}$, denoted as $\hat{\textbf{s}}$, is obtained as the output of the digital processing module $\textbf{B}$, yielding
\begin{equation}
\hat{\textbf{s}}=\textbf{B}\tilde{\textbf{z}}.
\end{equation}
The problem now is to jointly design the analog combiner $\textbf{A}$, the dynamic range $\gamma$, and the digital processing matrix $\textbf{B}$, so that the mean square error (MSE) of the task estimate can be minimized. Mathematically, we have the following
objective
\begin{equation}
\min\limits_{\textbf{A},\gamma,\textbf{B}} E\left[||\textbf{s}-\hat{\textbf{s}}||^2\right].
\label{Eq:Objective}
\end{equation}
\subsection{Theoretical Results}
\label{subsec:TR}
According to the orthogonality principle, the MSE in \eqref{Eq:Objective},  $E[||{\textbf{s}}-\hat{\textbf{s}}||^2]$, can be re-expressed as
\begin{equation}
 E\left[||\textbf{s}-\hat{\textbf{s}}||^2\right]=
 E[||{\textbf{s}}-\tilde{\textbf{s}}||^2]+E[||\tilde{\textbf{s}}-\hat{\textbf{s}}||^2],
\end{equation}
where $\tilde{\textbf{s}}$ is the linear minimum mean square error (LMMSE) estimate of $\textbf{s}$ from $\textbf{y}$, that is,
$\tilde{\textbf{s}}=\boldsymbol{\Gamma}\textbf{y}$, with $\boldsymbol{\Gamma}$ denoting the LMMSE estimation matrix. Note that the first term in the above equation is independent of $\hat{\textbf{s}}$. The optimization problem in \eqref{Eq:Objective} can thus be equivalently replaced by
\begin{equation}
\min\limits_{\textbf{A},\gamma,\textbf{B}} E\left[||\tilde{\textbf{s}}-\hat{\textbf{s}}||^2\right],
\label{Eq:Objective2}
\end{equation}
which is the same as \cite{Shlezinger_Tsp19}. Therefore, in the following, we directly provide the obtained optimization results and omit the proof.

Let $\boldsymbol\Sigma_\textbf{y}$ be the covariance matrix of the received signal $\textbf{y}$, and $w_l, \,l=1,\ldots,K$ the dither signal added to the input of the $l$th quantizer.
Let $\textbf{A}^{\circ}$ and $\textbf{B}^{\circ}$  the optimal analog and digital processing matrices that achieve the minimal MSE distortion. Then
we have the following results
\cite{Shlezinger_Tsp19}:

\textit{Theorem 1:}
For any analog combining matrix $\textbf{A}$ and dynamic range $\gamma$ such that $\text{Pr}{\left( |(\textbf{Ay})_l+w_l|>\gamma\right)=0}$, namely, the quantizers operate within their dynamic range with probability one, the digital processing matrix which minimizes the MSE is given by
\begin{equation}
\textbf{B}^{\circ}(\textbf{A})=\boldsymbol\Gamma \boldsymbol\Sigma_\textbf{y} \textbf{A}^H \left( \textbf{A} \boldsymbol\Sigma_\textbf{y} \textbf{A} ^H+\frac{2\gamma^2}{\tilde{M}_K^2\cdot K}\textbf{I}_K\right)^{-1}.
\label{Eq:B}
\end{equation}

\textit{Theorem 2:}
For the hardware-limited quantization system based on the model depicted in Fig. \ref{Fig1}, the optimal analog combining matrix is given by $\textbf{A}^{\circ}=\textbf{U}_{\textbf{A}}\boldsymbol\Lambda_{\textbf{A}}\textbf{V}_{\textbf{A}}^H\boldsymbol\Sigma_{\textbf{y}}^{-1/2}$, where
\begin{itemize}
\item $\textbf{V}_{\textbf{A}}\in \mathcal{C}^{N\times N}$ is the right singular vectors matrix of
$\tilde{\boldsymbol\Gamma}\triangleq \boldsymbol\Gamma \boldsymbol\Sigma_{\textbf{y}}^{1/2}$.
\item $\boldsymbol\Lambda_{\textbf{A}} \in \mathcal{C}^{K\times N}$ is a diagonal matrix with diagonal entries
      $$(\boldsymbol\Lambda_{\textbf{A}})_{i,i}^2=\frac{2\kappa_p}{\tilde{M}_K^2\cdot K} \left(\zeta \cdot \lambda_{\tilde{\boldsymbol\Gamma},i}-1\right)^{+}$$
      where $\kappa_p=\eta^2\left(1-\frac{2\eta^2}{3\tilde{M}_K^2} \right)^{-1}$ with $\eta$ denoting a constant that is set to guarantee that the quantizer operates within the dynamic range \cite{Shlezinger_Tsp19}, $\{\lambda_{\tilde{\boldsymbol\Gamma},i}\}$ are singular values of $\tilde{\boldsymbol\Gamma}$ arranged in a descending order,
      and $\zeta$ is chosen such that
      $$\frac{2\kappa_p}{\tilde{M}_K^2\cdot K}\sum_{i=1}^{K}\left(\zeta \cdot \lambda_{\tilde{\boldsymbol\Gamma},i}-1\right)^{+}=1.$$
\item $\textbf{U}_{\textbf{A}}\in \mathcal{C}^{K\times K}$ is a unitary matrix which guarantees that $\textbf{U}_{\textbf{A}}\boldsymbol\Lambda_{\textbf{A}}\boldsymbol\Lambda_{\textbf{A}}^H\textbf{U}_{\textbf{A}}^H$ has identical diagonal entries.
\end{itemize}
The dynamic range of the quantizer is given by
\begin{equation}
\gamma^2=\frac{\eta^2}{K}\left(1-\frac{2\eta^2}{3\tilde{M}_K^2} \right)^{-1},
\label{Eq:dy_range}
\end{equation}
and the resulting minimal achievable distortion is
\begin{equation}
E[||\tilde{\textbf{s}}-\hat{\textbf{s}}||^2]=
\sum_{i=1}^{K}\frac{\lambda_{\tilde{\boldsymbol\Gamma},i}^2}{\left(\zeta \cdot \lambda_{\tilde{\boldsymbol\Gamma},i}-1\right)^{+}+1}.
\end{equation}
In our prototype, we configure the analog combiner according to Theorem 2, and the dynamic range of the scalar quantizer based on
\eqref{Eq:dy_range}. The calculated matrix $\textbf{B}^{\circ}$ in \eqref{Eq:B} is used for the task vector recovery in the digital domain.
Details of the hardware implementation are discussed in the next section.
\section{Hardware Implementation}
\label{sec:hardware}
This section elaborates on the system architecture of the hardware prototype, which realizes task-based quantization for multi-user signal recovery detailed in the previous section. We first present the high-level system architecture in subsection \ref{subsec:4.1}. The concrete structure of each component is provided in subsection \ref{subsec:4.2}, and the design challenges are detailed in subsection \ref{subsec:4.3}.
\subsection{High-level Architecture}
\label{subsec:4.1}
%
%

\begin{table}[t!]
\centering
\caption{List of Hardware Components}
\label{table:hardware_components}
\begin{tabular}{||c c c||}
 \hline
Component & Model Number & Make \\
&& \\ \hline
FPGA & Xilinx VC707 & Texas Instruments \\
 && \\
DAC & FMC204 FPGA Card & Abaco Systems\\
&& \\
ADC  & FMC168 FPGA Card  & Abaco Systems \\
&& \\
Local oscillator& VSG25A & Signal Hound\\
  && \\ [1ex]
\hline
\end{tabular}
\end{table}
Fig. \ref{Analog_Precoder_system} shows our hardware board, which consists of five main blocks: GUI, Signal generator, Analog combiner board, Sampling, and Computing center. Details of the employed hardware components are presented in
Table \ref{table:hardware_components}, and the major building components are as follows:


\begin{figure}[t]
    \centering
    \includegraphics[width=0.48\textwidth]{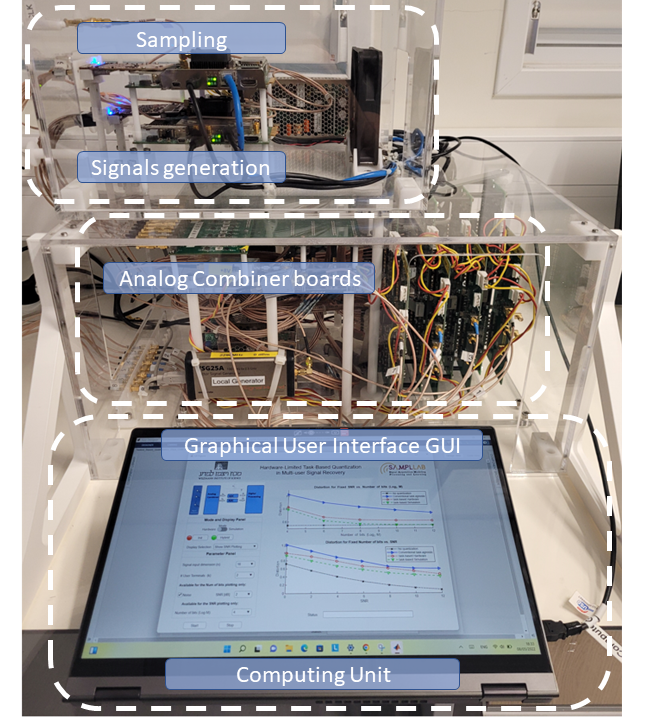}
    \caption{The task-based quantization system.}
    \label{Analog_Precoder_system}
\end{figure}

\begin{figure*}[!t]
	\centering
	\includegraphics[width=0.9\textwidth]{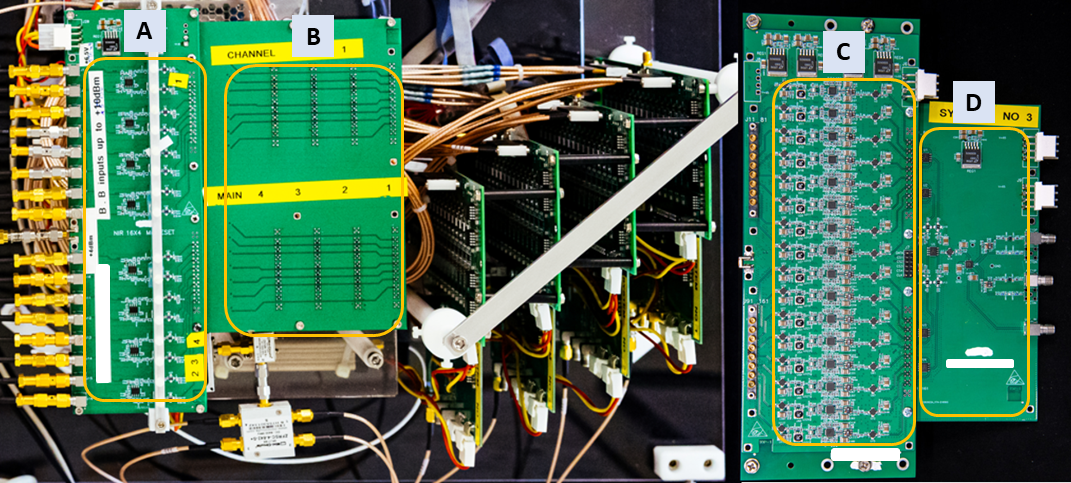}
	\caption{The analog combiner hardware board.}
	\label{Fig:hw_board}
\end{figure*}

\begin{figure}[h]
    \centering
    \includegraphics[width=(0.48\textwidth)]{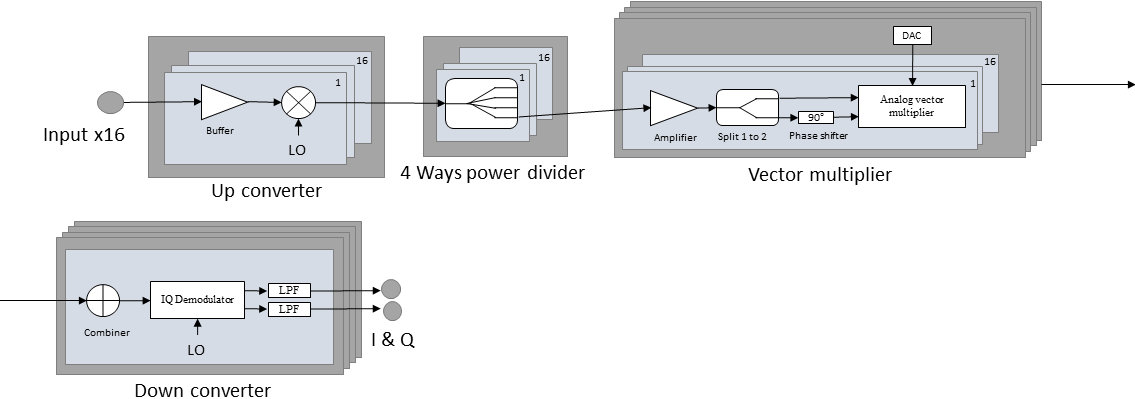}
    \caption{Analog combiner schematics.}
    \label{Analog_combiner_flow}
\end{figure}
%


\subsubsection{GUI}\label{Comp_GUI}
The graphical user interface (GUI) is used for controlling the system parameters, which allows the user to configure the experimental setup in a user-friendly environment. The main controllable parameters include the number of user terminals, receiving antennas, quantization bits, and the SNRs of the received signals.
Based on these parameters, the MATLAB running on the computing unit generates input data for 16-channel digital-to-analog converters (DACs) that are located at the analog combiner board  (details in \ref{AP_board}), and the optimal weights for the analog combiner configuration.

\subsubsection{Signal Generator}\label{Sig_gen}
The digital data generated by MATLAB is then fed to an Field Programmable Gate Array (FPGA) board with 16 transmit channels DACs to generate analog waveform signals. This process is adopted to mimic the real-world receiving signals at the base station (BS).

\subsubsection{Analog Combiner Board}\label{AP_board}
The 16 baseband analog signals from the DAC are next fed into the analog combiner board, as illustrated in Fig. \ref{Fig:hw_board}. To transmit them in the desired frequency, they are up-converted to 2.3GHz by 16 dedicated modulators. Then, the signal of each channel is passed through a 4-way power divider (splitter),  yielding 64 analog RF signals in total. The 64 signals are then fed into a 4-combiner boards.
Each combiner board is fed by 16 channels and has a single output.
The combiner board is controlled by an analog vector multipliers device designed to control a signal's gain and phase. The overall process is illustrated in Fig. \ref{Analog_combiner_flow}.
In this way, the tailored analog combiner board converts 16-channel signals to 4, implementing the function of the aforementioned analog combiner matrix $\textbf{A}$.

\subsubsection{Sampling}\label{Sampling}
The outputs (both I and Q) of the analog combiner board are fed into a sampling board. The four analog signals are down-converted from 2.3 GHz to 20 MHz, and are converted to digital signals by using 4DSP FMC168 16-bit digitizer card.

\subsubsection{Computing Center}\label{Computing_center}
The four digital streams are then transferred to the Matlab application on the computing center. The Matlab mimics a digital low-bit quantization and then recovers multi-user signals in the digital domain.
Finally, the results are displayed on the GUI to demonstrate the signal recovery performance of the task-based hardware prototype.




%

\subsection{Details of Each Block}
\label{subsec:4.2}
%
%
\subsubsection{Waveform generation}
The 16 digital baseband signals generated by the host application are transferred to the FPGA board in real-time by an Ethernet cable. The FPGA board generates the corresponding analog baseband signals waveform with a maximal frequency range of  100 MHz.
\subsubsection{Analog Combiner}
The analog combiner board is a self-designed dedicated hardware that realizes a controllable analog combiner network. As shown in Fig. \ref{Fig:hw_board}, the board consists of four parts:

A. Up-conversion: The 16 input complex baseband (BB) signals, whose maximal frequency range is 100MHz, are up-converted to RF signals using a 2.3 GHz carrier waveform. The carrier is generated by a VSG25A vector signal generator. By up-conversion, the RF signals can represent the passband signals observed at the base station.

B. Passband signals splitting: The analog passband signal of each channel is split into four. In the considered setting here, we have 64 analog RF signals in total, which are combined for further processing. The board can support 4 RF-chain processing. Since the number of users is set as 2 in this experiment, we only use 2 of them to  process the output RF signals.

C. Parameter generation and  configuration for the combiner: Each split signal is fed into an amplifier, split again into two signals with a 90-degree offset.  The two signals then enter into an ADL5390 analog vector multiplier.
The analog vector multiplier
implements the phase and gain of each analog combining weight,  which is applied to combine the input signal.
The applied weights are determined by the output DC level of an AD5674 octal 12-bit DACs with serial load capabilities, which receives control commands via Arduino Nano microcontroller device to configure the analog combining weights.
The usage of controllable gains and phases requires a calibration stage when the interconnections are established, to guarantee that the configured weights are correctly translated into the desired phase and gain values.

D. Summing up of the incoming signals and down-conversion:
The final step is summing the 16 output signals of each group after weighting, to obtain a combined passband signal. The signal is then down-converted by the same local oscillator that is employed for up-conversion, and filtered to baseband with a maximum 100 MHz bandwidth.

\subsubsection{Quantization}
The four output signals are forwarded to be sampled by the 4DSP FMC168 16-bit digitizer card. However, in task-based quantization, it is expected to use low-bit quantizers. We here use software simulation to mimic the hardware implementation of such a scalar quantizer defined as
\begin{equation}
q(x)\triangleq
\begin{cases}
\Delta \left(\lfloor\frac{x}{\Delta}\rfloor+\frac{1}{2}\right), & \text{for} \quad |x|<\gamma \\
\text{sign}(x)(\gamma-\frac{\Delta}{2}), & \text{else},
\end{cases}
\end{equation}
where $x$ is the input signal,
$\Delta=\frac{2\gamma}{\tilde{M}_K}$ represents the quantization spacing. The variable $\tilde{M}_K$ is varied in the experiments for different number of bits. The symbol
$\text{sign}(\cdot)$ denotes the signum function, given by
\begin{equation}
\text{sign}(x)\triangleq
\begin{cases}
+1, &  x\ge 0, \\
-1, & \text{else}.
\end{cases}
\end{equation}
\subsubsection{Software (digital processing)}
\begin{figure}[t]
    \centering
    \includegraphics[width=0.48\textwidth]{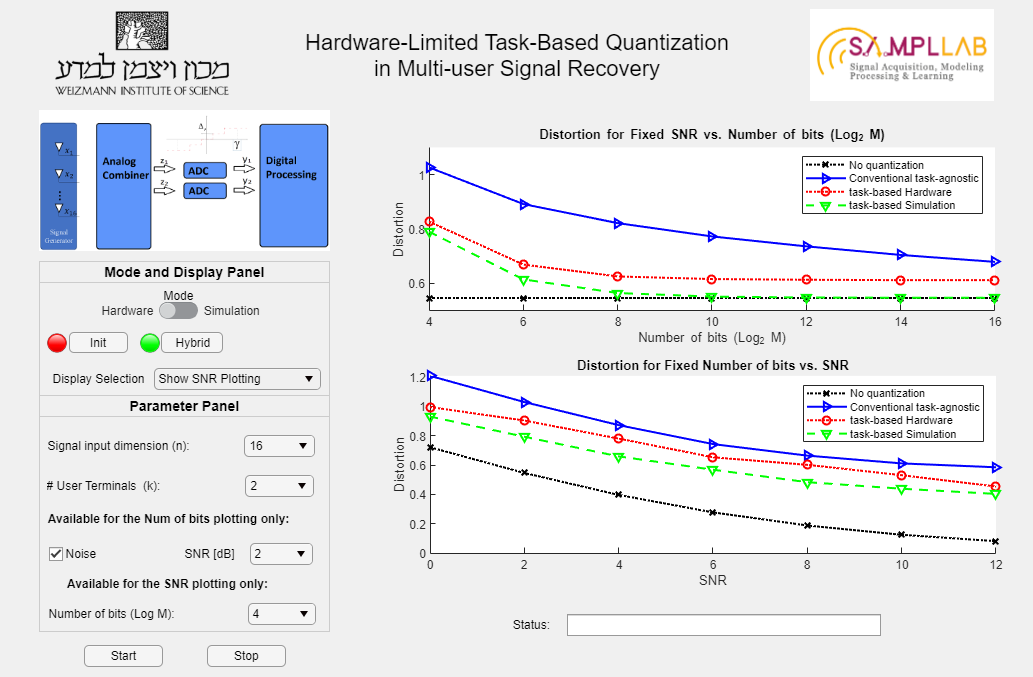}
    \caption{Overview of the GUI.}
    \label{Fig3}
\end{figure}
The software part consists of two components: a computing center running the MATLAB-based host application, and a GUI-based control and display interface.

The computing center is a 64-bit computer with 8 CPU cores and 16GB RAM running the MATLAB-based host application. The application is responsible for generating the digital baseband signal, computing the optimal analog and digital processing matrices as detailed in Theorems 1 and 2, computing the dynamic range of the quantizer, and post-processing the digital output to recover the task vector.

The display part of the GUI presents the experiment results in two  modes: the MSE distortion with respect to the number of bits, or SNR, as shown in Fig. \ref{Fig3}. The control part provides a way for users to interact with the experiment setup, that is, it allows users to change the parameters used in the experiment.
The main controllable parameters include the dimensionality of the received signal and the task vector, the SNR level for plotting MSE distortion versus the number of bits, and the number of bits for plotting MSE distortion versus SNR. Details of the supported parameter combinations are summarized in Table \ref{Tab:I}.

\begin{table*}[t!]
	\centering
	\caption{Controllable Parameters Supported by GUI}
	\label{Tab:I}
\setlength{\tabcolsep}{1 mm}{
		\scalebox{0.75}{
		\begin{tabular}{|c|c|c|c|c|}
	    \hline
         \rule{0pt}{10pt}  Working mode & \multicolumn{2}{c|}{Simulation} & \multicolumn{2}{c|}{Hardware}  \\
        \hline
        \rule{0pt}{10pt} Curve display mode & NMSE vs. Number of bits & NMSE vs. SNR & NMSE vs. Number of bits & NMSE vs. SNR\\
        \hline
        \rule{0pt}{10pt} Number of UTs & \multicolumn{2}{c|}{$K=2,4,8$} & \multicolumn{2}{c|}{$K=2$} \\
        \hline
        \rule{0pt}{10pt} Number of receiving antennas in the BS & \multicolumn{2}{c|}{$N=4,8,16,60,120$} & \multicolumn{2}{c|}{$N=16$} \\
       \hline
       \rule{0pt}{10pt} Noise & $\text{SNR}=2,4,6,8,10$ & &  $\text{SNR}=2,4,6,8,10$&\\
       \hline
       \rule{0pt}{10pt} Number of bits ($\text{Log}_2 M$) & &$4,8,12,16,20$ &  &$4,8,12,16,20$\\
      \hline
  \end{tabular} }
  }
\end{table*}
%
%

\begin{figure}[!h]
	\centering
	\begin{tabular}{c}
		\subfigure[]{\includegraphics[width=3.2in]{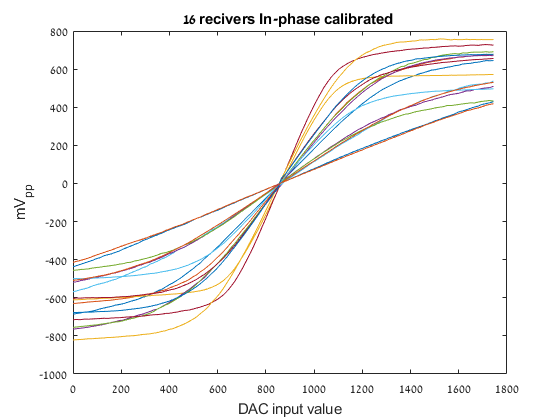}\label{fig:board1_calibration_I}} \\
		\subfigure[]{\includegraphics[width=3.2in]{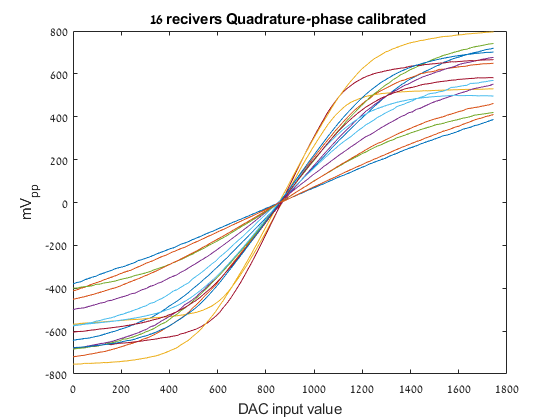}\label{fig:board1_calibration_Q}} \\
		\subfigure[]{\includegraphics[width=3.2in]{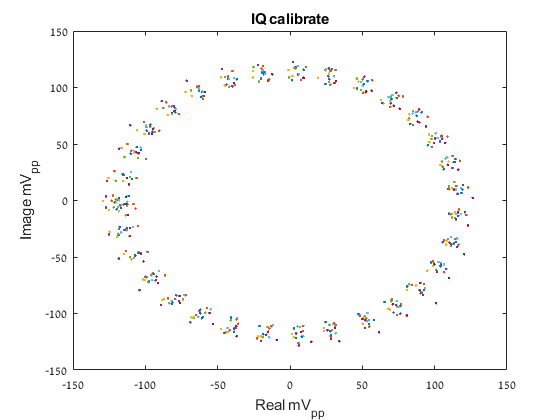}\label{fig:board1_calibration_verafication}}
	\end{tabular}
	\caption{Calibration process results.}
	\label{fig:board1_calibration}
\end{figure}

\subsection{Design Challenges}
\label{subsec:4.3}
One of the critical challenges in implementing the analog combiner board is to guarantee that all RF chains operate within the linear dynamic range of the device. This will ensure that the combination of all 16 channels for each of our four output boards will result in an accurate summation. In our case, there are $16\times4 = 64$ RF chains that need to be calibrated, and each of their amplitude and phases need to be adjusted. In order to overcome this challenge, we introduced a calibration process that scanned through the amplitude and phase of each RF-chain and performed relevant modifications. This process is done by setting the DAC value for adjusting the I and Q amplitude for each signal, as shown in Fig. \ref{fig:board1_calibration_I} and Fig. \ref{fig:board1_calibration_Q}.
Specifically,
Fig. \ref{fig:board1_calibration_I} presents the In-Phases signals which are received from the 16 channels in a single board, while Fig. \ref{fig:board1_calibration_Q} presents the Quadrature-Phases signals which are received from the 16 channels in the same board.
Fig. \ref{fig:board1_calibration_verafication} represents the calibrated 16 RF-chain signals from the output board.
The process is an iterative process that identifies the best linear point in which the Euclidean distance from the center is optimal.
%
%
\section{Hardware Results}
\label{sec:Experimental}
In this section, hardware experiments are carried out to evaluate the performance of task-based quantization in multi-user signal recovery. We consider the case where the number of users is $K=2$, and the number of antennas at the BS is $N=16$. The transmitted signal from the two users obeys zero-mean and unit variance Gaussian distribution, and the channel is generated based on \eqref{Eq: channel} with $L=3$ paths for each user. All the results are obtained by averaging 2000 experiments.
\begin{figure}[h]
    \centering
    \includegraphics[width=0.48\textwidth]{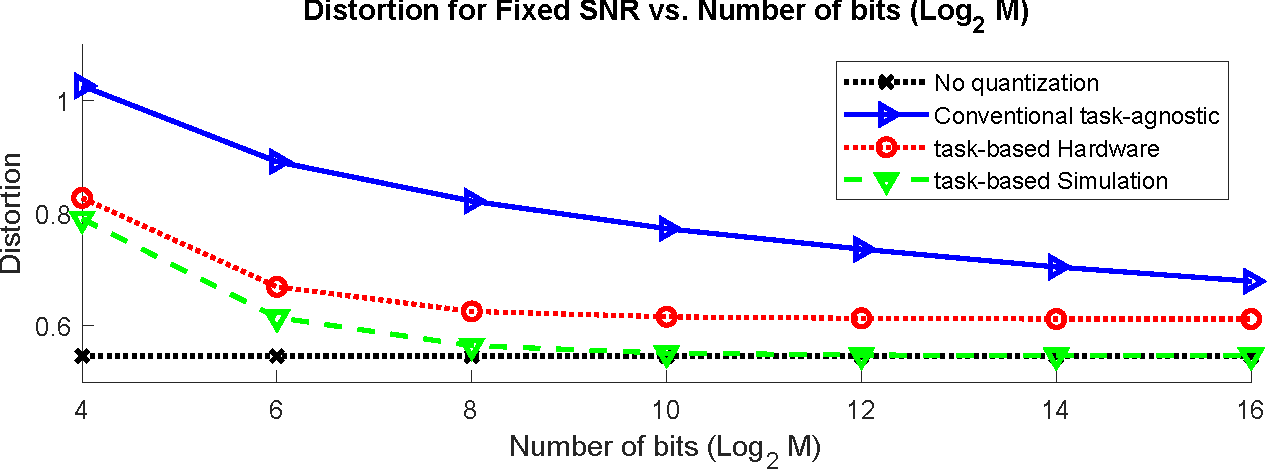}
    \caption{The MSE distortion versus the number of total bits.}
    \label{Fig4}
\end{figure}
\begin{figure}[h]
    \centering
    \includegraphics[width=0.48\textwidth]{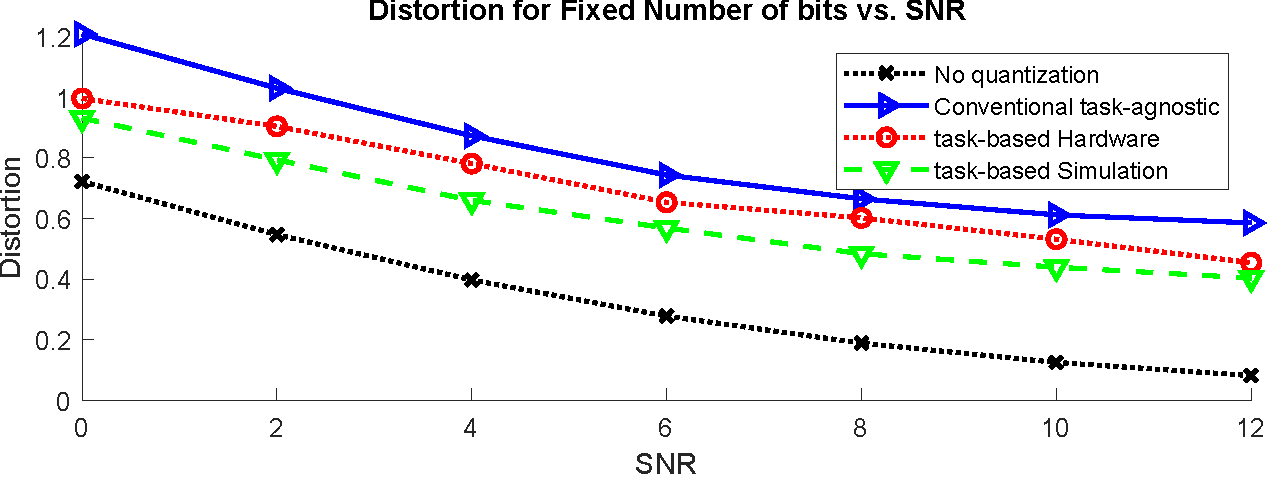}
    \caption{The MSE distortion versus the SNRs.}
    \label{Fig5}
\end{figure}

As a comparison, task-agnostic vector quantization results are included. Different from scalar quantizers which operate on a scalar input, vector quantizers have a multivariate input. Therefore, vector quantization cannot be implemented using practical serial scalar ADCs. Here, we employ simulated task-agnostic vector quantization as a comparison since it represents the best system one can construct when the quantizer is designed separately from the task \cite{Shlezinger_Tsp19}. Furthermore, the ideal case where no quantization is imposed on the sampled signal is also provided as a benchmark.
The GUI provides two modes: simulation mode and hardware mode.
In the simulation mode, the combining, sampling and quantization of the received signal are all performed by software, i.e., MATLAB. In hardware mode, the combining and sampling are performed by the hardware board.
We set $\text{SNR}=2$ dB in the case of displaying distortion versus the number of bits, and the number of total used equals $4$ for plotting distortion versus SNRs. The results are shown in Fig. \ref{Fig4} and Fig. \ref{Fig5} where both the simulation and hardware results are provided in the same figure.

From these results, we see that task-based quantization significantly outperforms task-agnostic vector quantization, and can approach the optimal performance with the increase of $M$. In particular, when each quantizer is assigned more than five bits, i.e., $\text{log}_2 M\ge 5K$, the quantization error becomes negligible. This suggests that by exploiting prior knowledge of the task, and by properly designing the overall system, task-based quantization can achieve satisfying performance with a much smaller number of bits, i.e., from $16$ receivers with $16$ bits each to $2$ receivers with $5$ bits each.
Furthermore, the hardware results agree with the simulated ones, with only a small performance gap caused by imperfect hardware calibration and hardware noise, verifying the effectiveness of the task-based quantization hardware prototype.
\section{Conclusion}
With the increase of data rate, conventional analog-to-digital converters (ADCs) which sample at the Nyquist rate and use high-resolution quantizers face challenges in storage and power consumption. To reduce quantization bits, task-based quantization has been proposed by exploiting the underlying task for the system design. In this work, we presented the application of task-based quantization in multi-user signal recovery and provided a hardware implementation. The prototype consists of a tailored configurable analog combiner board and a software-based processing and demonstration system. Experimental results illustrate the superiority of task-based quantization over conventional ADCs, mitigating the gap between the theory and its practical application.
\label{sec:Conclusions}

\small
\bibliographystyle{IEEEbib}
\bibliography{yourbibliography}
			
\end{document}